\documentclass[
aps,
prl,
reprint,
notitlepage,
superscriptaddress,
twocolumn,
floatfix,
amssymb,
nofootinbib,
]{revtex4-2}
\usepackage[utf8]{inputenc}
\usepackage[english]{babel}
\usepackage{array}
\usepackage{amsmath}
\usepackage{amssymb}
\usepackage{graphicx}
\usepackage{blochsphere}
\usepackage{braket}
\usepackage{color}
\usepackage{float}
\usepackage{soul}
\usepackage{enumitem}
\usepackage[dvipsnames]{xcolor}
\usepackage{comment}
\usepackage{siunitx}
\DeclareSIUnit{\gauss}{G}
\DeclareMathOperator\Arg{Arg}
\usepackage{wrapfig}
\usepackage{needspace}
\usepackage{orcidlink}

\usepackage{appendix}

\usepackage{hyperref}
\hypersetup{
		unicode=true,
		linktoc=all,
		verbose=true,
		final=true,
		plainpages=false,
		pdfstartview=FitV,
		pdftoolbar=true,
		pdfmenubar=true,
		pdfpagelayout=OneColumn,
        bookmarksopen=true,
		bookmarksnumbered=true,
		breaklinks=true,
		linktocpage=true,
		colorlinks=true,
		linkcolor=blue,
		urlcolor=blue,
		citecolor=blue,
		anchorcolor=green
	}

\expandafter\let\csname equation*\endcsname\relax
\expandafter\let\csname endequation*\endcsname\relax
\expandafter\let\csname eqnarray*\endcsname\relax
\expandafter\let\csname endeqnarray*\endcsname\relax

\newcommand{\A}{\nonumber}
\newcommand{\RN}[1]{\textup{\uppercase\expandafter{\romannumeral#1}}}

\makeatletter
\newcommand{\fmarki}{*}
\newcommand{\fmarkii}{\ensuremath{\dagger}}
\newcommand{\fmarkiii}{\ensuremath{\ddagger}}
\newcommand{\fmarkiv}{\ensuremath{\mathsection}}
\newcommand{\fmarkv}{\ensuremath{\mathparagraph}}
\newcommand{\fmarkvi}{\ensuremath{\|}}
\newcommand{\fmarkvii}{**}
\newcommand{\fmarkviii}{\ensuremath{\dagger\dagger}}
\newcommand{\fmarkix}{\ensuremath{\ddagger\ddagger}}
                
\def\@fnsymbol#1{{\ifcase#1\or \fmarki\or \fmarkii\or \fmarkiii\or \fmarkiv\or \fmarkv\or \fmarkvi\or \fmarkvii\or \fmarkviii\or \fmarkix \else\@ctrerr\fi}}
\makeatother

\renewcommand{\fmarki}{+}
\renewcommand{\fmarkii}{\textdagger}
\renewcommand{\fmarkiii}{\ensuremath{\mathparagraph}}

\usepackage{silence}
\usepackage{tikz}
\usepackage[siunitx, european, oldvoltagedirection]{circuitikz}	
\usetikzlibrary{positioning, arrows, shapes, calc, 3d, math, quotes, angles, tikzmark}
\tikzset{
every picture/.style={/utils/exec={\sffamily}},
every node/.style={font=\scriptsize},
}

\usepackage{environ}
\makeatletter
\newsavebox{\measure@tikzpicture}
\NewEnviron{scaletikzpicturetowidth}[1]{%
	\def\tikz@width{#1}%
	\begin{lrbox}{\measure@tikzpicture}%
		\BODY
	\end{lrbox}%
	\pgfmathparse{#1/\wd\measure@tikzpicture}%
	\BODY
}
\makeatother

\def\beamwidth{1.8pt}
\def\widthscale{0.023}
\tikzset{
	SEMBeam/.style = {line width=\beamwidth, eBeamColor},
	dot/.style = {circle, fill, minimum size=#1,
		inner sep=0pt, outer sep=0pt},
	dot/.default = 6pt,  
    line/.style = {draw, -{latex}},
    eBeamField/.style = {dashed, B1Color, very thick},
	shadowstyle/.style = {line width = 0.5*\beamwidth},
	shadowed/.style={ 
		preaction={transform canvas = {shift = {(0pt,1*\beamwidth*\widthscale,0pt)}}, draw=eBeamColor!75, shadowstyle},
		preaction={transform canvas = {shift = {(0pt,2*\beamwidth*\widthscale,0pt)}}, draw=eBeamColor!55, shadowstyle},
		preaction={transform canvas = {shift = {(0pt,3*\beamwidth*\widthscale,0pt)}}, draw=eBeamColor!35, shadowstyle},
		preaction={transform canvas = {shift = {(0pt,4*\beamwidth*\widthscale,0pt)}}, draw=eBeamColor!15, shadowstyle}},
        nosep/.style={inner sep=0pt, outer sep=0pt},
        zoomline/.style = {dashed, -}
}

\tikzset{RFAmp/.style={
		muxdemux,
		muxdemux def={NL=1, Lh=3, NR=1, Rh=0, NB=0, NT=0},
        scale=0.6}
}

\tikzstyle{charge-}=[very thin,draw=black,top color=ElectronColor!50,bottom color=ElectronColor!80,shading angle=20,circle,inner sep=0.5]

\newcommand{\Modulation}[1]{
}

\newcommand{\Dircoupler}[1]{
   \begin{tikzpicture}[baseline=0ex, scale=#1]
        \draw[thick] (-0.5,0.5) -- (0.5,0.5);
        \draw[thick] (-0.35,0.65) -- (0,0.65) -- (0,1);
	\end{tikzpicture}
}

\usepackage{pgfplots}
\pgfplotsset{compat=1.10}

\pgfmathdeclarefunction{absorbtive}{3}{%
	\pgfmathparse{#3*2*(x-#1)/(#2*sqrt(2*pi)*(2*#2^2))*exp(-((x-#1)^2)/(2*#2^2))}%
}
\pgfmathdeclarefunction{dispersive}{3}{%
	\pgfmathparse{#3/(#2*sqrt(2*pi))*exp(-((x-#1)^2)/(2*#2^2))}%
}

\usepackage{xcolor}
\definecolor{eBeamColor}{HTML}{2B2D7C}

\definecolor{bdpaColor}{HTML}{FF2C00}
\definecolor{electronics0Color}{HTML}{FF9500}
\definecolor{electronics1Color}{HTML}{00B945}
\definecolor{electronics2Color}{HTML}{85d6B2}
\definecolor{electronics3Color}{HTML}{845B97}
\definecolor{electronics4Color}{HTML}{3F9EBD}

\definecolor{greyColor1}{HTML}{474747}
\definecolor{greyColor2}{HTML}{9e9e9e}
\definecolor{greyColor3}{HTML}{d3d3d3}

\definecolor{IChannelColor}{HTML}{1f77b4}
\definecolor{QChannelColor}{HTML}{ff7f0e}

\definecolor{steelColor}{HTML}{989898}
\definecolor{SEMColor}{HTML}{f4f0ec}
\definecolor{PhilipsColor}{HTML}{0247fe}
\definecolor{plotColorSignalDis}{HTML}{1f77b4}
\definecolor{plotColorSignalAbs}{HTML}{ff7f0e}

\colorlet{SpinColor}{electronics3Color}
\colorlet{biasFieldColor}{electronics0Color}
\colorlet{B1Color}{electronics4Color}
\colorlet{modulationColor}{electronics2Color!200}
\colorlet{levelColor}{electronics2Color}
\colorlet{deflectionColor}{greyColor1}
\colorlet{transitionColor}{bdpaColor}
\colorlet{semLensColor}{electronics0Color}
\colorlet{circuitColor}{greyColor1}
\colorlet{plotColorZeeman}{electronics1Color}
\colorlet{insertColor}{bdpaColor}
\colorlet{ElectronColor}{electronics1Color}
\colorlet{beamShadeColor}{eBeamColor!15}
\colorlet{microCoilColor}{greyColor1}
\colorlet{signalPlotColor}{electronics3Color}
\begin{document}
\setlength{\parindent}{0mm}

\title{Coherent Driving of a Quantum System with Modulated Free-Space Electrons}

\author{Matthias Kolb\orcidlink{0009-0007-4106-1191}\textsuperscript{*}}
\email{matthias.kolb@tuwien.ac.at}
\author{Thomas Spielauer\orcidlink{0009-0002-0979-7880}\textsuperscript{*}}
\email{thomas.spielauer@tuwien.ac.at}
\author{Thomas Weigner\orcidlink{0000-0003-4941-4995}\textsuperscript{*}}
\email{thomas.weigner@tuwien.ac.at}
\affiliation{Vienna Center for Quantum Science and Technology, Atominstitut, Technische Universität Wien, Stadionallee 2, 1020 Vienna, Austria}

\author{Giovanni Boero\orcidlink{0000-0003-4913-3114}}
\affiliation{Microsystems Laboratory, Ecole Polytechnique Fédérale de Lausanne (EPFL), 1015 Lausanne, Switzerland}
\affiliation{Center for Quantum Science and Engineering, Ecole Polytechnique Fédérale de Lausanne (EPFL), 1015 Lausanne, Switzerland}

\author{Dennis Rätzel\orcidlink{0000-0003-3452-6222}}
\affiliation{Vienna Center for Quantum Science and Technology, Atominstitut, Technische Universität Wien, Stadionallee 2, 1020 Vienna, Austria}
\affiliation{ZARM, University of Bremen, Am
Fallturm 2, 28359 Bremen,
Germany}
\affiliation{University Service Centre for Transmission Electron Microscopy, Technische Universität Wien, Stadionallee 2, 1020 Vienna, Austria
\phantomsection
\label{link:ustem}\\
\vspace{0.2cm}
(\textsuperscript{*} These authors contributed equally to this work.)}

\author{Philipp Haslinger\orcidlink{0000-0002-2911-4787}}
\affiliation{Vienna Center for Quantum Science and Technology, Atominstitut, Technische Universität Wien, Stadionallee 2, 1020 Vienna, Austria}
\affiliation{University Service Centre for Transmission Electron Microscopy, Technische Universität Wien, Stadionallee 2, 1020 Vienna, Austria
\phantomsection
\label{link:ustem}\\
\vspace{0.2cm}
(\textsuperscript{*} These authors contributed equally to this work.)}

\begin{abstract}
\hspace{-1em}Control of quantum systems typically relies on the interaction with electromagnetic radiation. In this study, we experimentally show that the electromagnetic near-field of a spatially modulated free-space electron beam can be used to drive spin systems, demonstrating free-electron-bound-electron resonant interaction. 
By periodically deflecting the electron beam of a scanning electron microscope in close proximity to a spin-active solid-state sample, and sweeping the deflection frequency across the spin resonance, we directly observe phase coherent coupling between the electron beam's near-field and the two spin states. This method relies only on classically shaping the electron beams transversal correlations and has the potential to enable coherent control of quantum systems with unprecedented, electron microscopic resolution, opening novel possibilities for advanced spectroscopic tools in nanotechnology.
\end{abstract}

\maketitle
\section*{Introduction}
\label{sec:introduction}
Coherent manipulation of quantum systems with electromagnetic radiation is widely used in quantum science from quantum information processing \cite{HAFFNER2008155,graham2022multi, karliPassiveDemultiplexedTwophoton2025} and quantum communication \cite{QuantumRepeatersQuantum, bhaskar2020experimental} to the search for new physics \cite{peikNuclearClocksTesting2021, budker2023,jackson_kimball_probing_2023,Hamilton2015Atom-interferometry}. 
Typically, the method's inherent limitation of spatial resolution is on the scale of the employed wavelength and can be overcome by elaborate techniques \cite{ nagpal2009ultrasmooth, davis2020ultrafast,betzig2006imaging, yuan2019detecting}.
Electron microscopy, on the other hand, allows for subatomic spatial resolution \cite{Ishikawa2023} and temporal precision down to sub-picosecond \cite{Lobastov2005, Zewail2006} and even atto-seconds scale \cite{kozakPonderomotiveGenerationDetection2018a, nabben2023attosecond, hui2024attosecond}.
It has been theoretically shown that quantum systems can coherently interact with a stream of electrons with temporally shaped wave functions \cite{Feist2015Quantum, kealhofer2016all, schonenberger2019generation, wang2020coherent} in the optical frequency range \cite{Gover2020free, Favro1971en, zhao2021quantum, ruimy2021toward, garcia2021optical,yalunin2021tailored, morimoto2021coherent} and electron beams that are temporally or spatially modulated in the microwave frequency range \cite{ratzel2021controlling}.
While the experimental realization of this concept has remained challenging, recent experiments have taken important steps toward its implementation \cite{ruimyFreeelectronQuantumOptics2025a,grzesik_quantum_2025}.

In this manuscript we report on the first demonstration of driving a quantum transition, a magnetic dipole transition of a spin system, phase-coherently using the spatially modulated electron beam of a scanning electron microscope \cite{ratzel2021controlling}. The driving mechanism relies on transversal correlations imprinted on the free electrons through electrostatic modulation by "classically" driven deflection plates in the microwave range. This works even if the individual electron wave packets are much shorter than the wavelength of the beam modulation\cite{ratzel2021controlling}.

\section*{Experimental Setup}
\begin{figure}[htbp]
\input{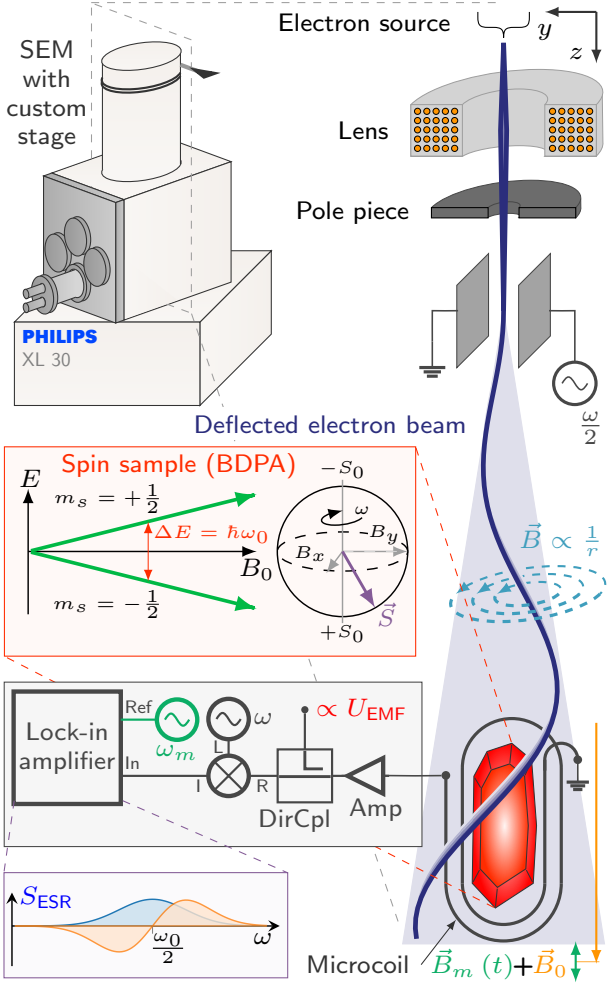}
\caption{Schematic view of our adapted Philips XL30 SEM that contains a miniaturized electron spin resonance (ESR) setup. It consists of an electron spin active sample (BDPA) surrounded by a microcoil with two windings, placed inside a coil (not shown) generating a bias field $B_0$ in $z$-direction, leading to Zeeman splitting of the spins in the sample. The SEM's electron beam focused by magnetic lenses is spatially modulated perpendicular to the $z-$direction by electrostatic deflection plates at a frequency $\omega/2 \sim 2\pi\times\SI{175}{\MHz}$. The second harmonic of the near-field of the electron beam (see Fig.~\ref{fig:sketch_harmonics_horizontal}) phase coherently drives the spins leading to precession in the $xy$ plane as depicted with $\vec{S}$ on the Bloch-sphere ($S_0$ being the thermal polarization). 
The magnetization from this spin precession inductively couples to the microcoil. The induced signal is amplified (Amp), while the induced voltage in the coil $U_\text{EMF}$ is measured directly via a directional coupler (DirCpl), the ESR part (R) is mixed down with an Local oscillator (L) to an intermediate signal (I). We measure it with a lock-in amplifier by modulating the bias field with ${B}_m\left(t\right) = A_m \cos\left(\omega_m t\right)$ ($\omega_m = 2\pi\times \SI{1.28}{\kHz}$). Scanning $\omega$ around half the resonance frequency $\omega_0 \approx 2\pi\times \SI{350}{\MHz}$ of the spin ensemble allows us to measure either a signal in phase with the driving field (usually called absorptive signal, orange) or one with \SI{90}{\degree} phase shift (usually called dispersive signal, blue) depending on the phase relation to the local oscillator.}
\label{fig:experimentSetup}
\end{figure}

The experimental setup consists of a modified Philips~XL30 Scanning Electron Microscope (SEM) equipped with a custom-made electrostatic beam modulator for spatially deflecting the electron beam and an electron spin resonance setup to inductively sense spin precessions induced by the modulated electron beam (see Fig.~\ref{fig:experimentSetup}).

The motorized sample stage of the SEM was replaced by a custom stage, comprised of an impedance-matched microcoil (two windings with an inner area of $1.1\times 0.6\,\si{\mm^2}$) containing \SI{\sim 0.5}{mm^3} of the electron-spin active radical $\alpha$,$\gamma$-Bisdiphenylen-$\beta$-phenylallyl (BDPA) at room temperature and a coil to generate the bias magnetic field $B_0 \approx \SI{12.5}{\milli \tesla}$ parallel to the propagation direction of the electron beam. This field leads to a spin polarization of the sample of about $\num{2.3e-5}$. The bias field $B_0$ separates the two states of each individual spin by \SI{\sim 28}{\GHz\per\T} leading to a tunable resonance frequency $\omega_0$ around $2\pi\times \SI{350}{\MHz}$.

The electron beam creates a magnetic field ${B}$, perpendicular to ${B}_0$. As shown in Fig.~\ref{fig:experimentSetup}, the electrons are manipulated by electrostatic deflection plates (with a deflection angle \SI{< 1}{\degree}) after traveling through the pole piece of the scanning electron microscope. Subsequently, the spatially (parallel to the microcoil) modulated electron beam passes the BDPA sample in close proximity. A modulation frequency of $\frac{\omega}{2} \approx 2\pi\times\SI{175}{\MHz}$ imprinted on the electron beam leads to oscillations of its position-dependent non-radiative magnetic near-field at higher harmonics, including the second harmonic close to the resonance frequency $\omega_0$. This approach allows us to suppress parasitic excitations of the spin ensemble by leaking microwaves e.g. from the electrostatic beam modulator.

In our experiment with a beam current in a range of $\SIrange[range-phrase = -, range-units = single]{\sim 2.5}{5.5}{\micro \ampere}$ at \SI{10}{\kilo \electronvolt}, at a distance of $h=\SI{1}{\milli \meter}$ measured from the top of the spin sample, with a deflection amplitude $A=\SI{0.9}{\mm}$ (see Fig.~\ref{fig:sketch_harmonics_horizontal}\textbf{a}) the resulting magnetic near-field amplitude of the second harmonic Fourier component is $B_1 \approx \SI{50}{\pico\T\per\micro\A}$ (see Section~\ref{sec:simulations}). This current corresponds to on average just $\numrange[range-phrase = -, range-units = single]{\sim 250}{550}{}$ beam electrons within a \SI{1}{\mm} vicinity of the sample. Although the electrons are Poisson-distributed, the modulation linewidth of their near-field does not depend on the current \cite{ratzel2021controlling} and remains unaffected.

Due to the small $B_1$ we perform ESR in the steady-state.
The electron beam's magnetic near-field $B$ excites the spin ensemble in the sample phase-coherently and consequently leads to a precession in the $xy$ plane, which inductively couples to the microcoil. 
The induced voltage is guided out of the SEM, frequency filtered, amplified, mixed down and recorded by a low-noise lock-in amplifier with a bandwidth of $\approx \SI{0.5}{\Hz}$ (see Fig.~\ref{fig:experimentSetup} purple box). For lock-in detection we additionally modulate $B_0$ with a field $B_m$ at $\omega_m \approx 2\pi\times \SI{1.28}{\kHz}$ with a peak amplitude $A_m$ of $\approx \SI{18}{\micro \T}$.

\section*{High Harmonic Generation}
\label{sec:higherHarmonic}
\begin{figure*}[htb]
    \centering
    \includegraphics{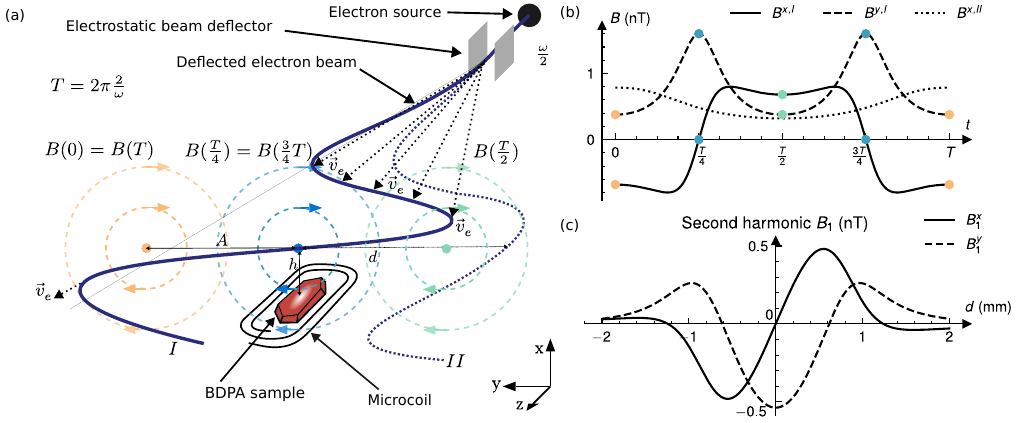}
    \caption{The electron beam, moving in $z$-direction, is deflected along the $y$-axis. This spatial modulation lets single electrons at velocity $\vec{v}_e$ travel on straight trajectories under changing angles. They form a sinusoidal shaped beam with a peak amplitude $A$ at a distance $h$ above ($+x$ direction) the spin sample (BDPA), placed within the microcoil. The resulting magnetic field lines are sketched for three different times of the modulation period in \textbf{a}, the corresponding time-dependent magnetic field is depicted in \textbf{b}. Since the $B^{y,I}$ field component consistently points in positive direction, a strong second harmonic component (double frequency) is created. When the entire beam is moving on one side of the sample ($d>A$, dashed beam $II$ in \textbf{a}) a \SI{180}{\degree} phase jump is observed ($B_{x, II}$, compared to $B_{x, I}$ as depicted in \textbf{b}). 
    Figure~\textbf{c} shows the amplitudes of the second harmonic Fourier component of the magnetic field in $x$ and $y$ direction for different beam positions $d$ along the $y$-axis measured from the coil center $d=0$, at a deflection amplitude of $A=\SI{0.9}{\mm}$ at $h=\SI{0.5}{\mm}$.}
    \label{fig:sketch_harmonics_horizontal}
\end{figure*}

Fig.~\ref{fig:sketch_harmonics_horizontal}\textbf{a} illustrates the non-radiative magnetic near-field of an electron beam moving in $z$-direction, deflected along the $y$-axis (parallel to the microcoil). The deflection creates a spatial modulation of the beam path in the $yz$ plane. Note, each electron moves along a straight line and is not accelerated after passing the deflection plates. The beam modulation leads to a temporal variation of the position-dependent magnetic near-field at the sample. 
Due to the non-linear distance dependence of the magnetic field, the beam modulation creates higher harmonics of the base frequency $\omega/2$.
The magnetic field $B(t)$ at the same position along the beam-axis is drawn for different times $t$ (orange, blue and green). 

Fig.~\ref{fig:sketch_harmonics_horizontal}\textbf{b} shows the time varying magnetic field at the position of the spin sample. The magnetic field at times $0,\dots T$ is marked with the corresponding colored dots.
For a beam deflected around the beam-axis ($I$, solid blue line in Fig.~\ref{fig:sketch_harmonics_horizontal}\textbf{a}), the field $B^{x, I}$ vanishes in the center, but reaches its maximum magnitude symmetrically around the $z$-axis
with a sign change across the center.
In contrast, $B^{y,I}$ is continuously positive, decreasing toward the deflection maxima and increasing toward the center. This behavior leads to a strong second-harmonic component (double frequency) as the beam passes the center twice per cycle.
For a beam deflection centered around the spin sample ($d=0$), the $x$ and $y$ compoments of $B$ only include odd and even harmonics, respectively. In Fig.~\ref{fig:sketch_harmonics_horizontal}\textbf{c} the second harmonic frequency components $B_1^x$ and $B_1^y$ are plotted for different $d$ evaluated at a fixed height of $h=\SI{0.5}{\mm}$. The zero-crossing of $B_1^x$ can be explained as a \SI{180}{\degree} phase jump when the beam is no longer moving directly above the sample, which is exemplified by the dashed beam ($II$) in Fig.~\ref{fig:sketch_harmonics_horizontal}\textbf{a}, \textbf{b}.

\section*{Results}
\label{sec:Results}
\begin{figure*}
    \includegraphics{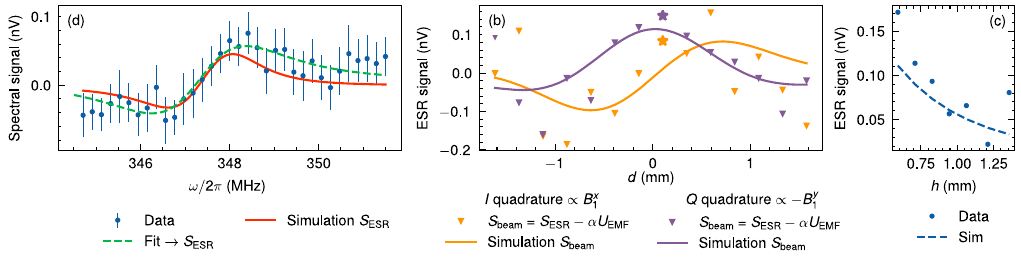}
    \caption{\textbf{a} ESR spectrum induced by a modulated electron beam at 
    $h\approx\SI{0.6}{\mm}$, $d\approx\SI{+0.1}{\mm}$, 
    with a beam deflection amplitude $A \approx \SI{0.9}{\mm}$
    (error bars represent the statistical standard error).
    The data were acquired using a differential measurement technique to suppress parasitic signals.
    The resulting spectrum is fitted using Equ.~\ref{equ:esrfit} and shows a clear absorptive signal at $\omega \approx 2\pi\times \SI{347.4}{\MHz}$, which aligns with conventional ESR data (see Appendix~\ref{sec:pcb}  Fig.~\ref{fig:bdpacrystal}\textbf{d} and  \textbf{e}). \textbf{b} shows the spatial dependence of the respective signal quadratures $I$ and $Q$ of the difference signal, determined from the extracted magnitudes and phases from the fits of individual ESR measurements
    (the position of the measurement in \textbf{a} is highlighted by the asterisk in \textbf{b}).
    Indirect driving of the spin system caused by the electromotive force induced by the beam into the microcoil, which creates a current and in turn a magnetic field that drives the spins was estimated from the measured induction (see Appendix~\ref{sec:main_IQAppendix} for details) and subtracted.
    The two quadratures in \textbf{b} show the direct driving $S_\text{beam}$ of the spins by the magnetic field with amplitudes $B_1^x$ and $B_1^y$ at frequency $\omega$ which excites the spins. The resulting magnetization is oscillating at frequency $\omega$ around the vertical $-z$-axis with \SI{90}{\degree} relative phase difference and induce the signal in \textbf{b}.
    The data show good agreement with the simulation (orange and purple solid lines).
    The simulation in \textbf{a} includes both the direct and indirect driving and compares well with the data.
    \textbf{c} depicts the spatial dependence of the beam driven ESR signal for different distances $h$ between beam and top of the sample. The signal magnitude is shown for $d\approx\SI{0.1}{\mm}$. 
    The measured data points (dots) exhibit a similar trend compared with the simulation (dashed line).
    Since the beam current varied during acquisition, the measurements have been rescaled to correspond to the values that would have been obtained at a constant beam current of \SI{1}{\micro\ampere}.}
    \label{fig:dataDiffEsr} 
\end{figure*}

To demonstrate phase coherent driving of the spin ensemble by the near-field, as well as the spatial dependence of ESR excitation by the modulated electron beam, we recorded spectra at different beam positions $d$. The beam was deflected with an amplitude of $A=\SI{0.9}{\mm}$ at the sample position approximately along the $y$-axis.

Each spectrum was recorded both with the electron beam switched on and off (averaged 25 times and integrated for \SI{600}{\ms} per measurement), enabling a differential measurement in order to suppress parasitic background signals (see Appendix~\ref{sec:diffmeasurement}).
The resulting spectra are fitted using Equ.~\ref{equ:esrfit} to extract magnitude (from the peak-to-peak voltage) and phase.
In Fig.~\ref{fig:dataDiffEsr}\text{a} the spectrum at beam position $h\approx\SI{0.6}{\mm}, d\approx\SI{0.1}{\mm}$, is shown. 
The fit yielded a resonance frequency of $\omega_0\approx2\pi\times\SI{347.4}{\MHz}$, and a linewidth of $\Delta\omega_{pp} \approx 2\pi\times\SI{2.1}{\MHz}$. The latter corresponds to a relaxation time $T_2 \approx\SI{87}{\ns}$ (see Appendices~\ref{sec:main_IQAppendix},~\ref{sec:pcb}).

This differential ESR signal consists of two contributions: the direct excitation from the near-field of the modulated electron beam ($S_\text{beam} \propto B_1^x - \mathrm{i}B_1^y$) and the indirect excitation via an electromotive force (EMF) induced in the microcoil. The EMF $U_\text{EMF}$, caused only by the $B_1^x$ component of the second harmonic near-field (see Fig.~\ref{fig:sketch_harmonics_horizontal}\textbf{d}), drives a current in the coil, limited by twice its resistance $2R_c$, which in turn creates a magnetic field that drives the spins, resulting in a signal component $S_\text{EMF} \propto U_\text{EMF}/(2R_c) \propto\mathrm{i}B_1^x$. By measuring $U_\text{EMF}$ we estimate the indirect driving signal $S_\text{EMF}$ using a scaling factor $\alpha$ to recover the direct driving signal $S_\text{beam}$ by simple subtraction (see Appendix~\ref{sec:coilexcitation}). The simulation shown in Fig.~\ref{fig:dataDiffEsr}\textbf{a} includes both effects.

Fig.~\ref{fig:dataDiffEsr}\textbf{b} compares the recovered signal $S_\text{beam}$ at $h\approx\SI{0.6}{\mm}$ for different $d$ with simulations that approximate the sample size.
The two quadratures represent the magnetization components, resulting from the driving $B_1^x$ and $-B_1^y$ field amplitudes, which are rotating around the negative $z$-axis at the resonance frequency $\omega_0$.
The results show a good qualitative agreement compared to the magnetic field shown in Fig.~\ref{fig:sketch_harmonics_horizontal}\textbf{d},
as well as a good quantitative agreement with the simulations.
While the latter includes several assumptions about the geometry (for details see Appendix~\ref{sec:simulations}), the former requires only a single scaling factor $\alpha$ and the phase alignment of the two signals $S_\text{ESR}$ and $S_\text{EMF}$ (see Appendix~\ref{sec:main_IQAppendix}).

Fig.~\ref{fig:dataDiffEsr}\textbf{c} illustrates the spatial dependence of the beam driven ESR signal $S_\text{beam}$ along the $x$-axis in comparison to the simulation (dashes). We represent the data using the signal magnitude rather than the $I$ quadrature. At $d\approx\SI{0.1}{\mm}$, where $B_1^x$ is close to zero, the response is thus dominated by the direct driving due to the $B_1^y$ field component.

\section*{Conclusions}
\label{sec:conclusion}

We demonstrated that the weak magnetic near-field generated by a modulated electron beam, consisting of only a few hundred electrons within a millimeter of the sample, can be harnessed through phase-coherent build-up to perform spin spectroscopy, in agreement with numerical simulations.

More than a century ago the Franck--Hertz experiment \cite{Franck1914Ueber} demonstrated the quantization of atomic energy levels via scattering of individual electrons. 
Ever since, excitation by electron scattering has only been realized by single electron events.
In contrast our approach demonstrates that a consecutive, coherent and frequency dependent build-up of excitation can be realized by an ensemble of collectively modulated but independent electrons.

The experimentally demonstrated effect, therefore, establishes a novel pathway for coherent control of quantum systems, harnessing the unique capabilities of the electron microscope platform \cite{Reimer2008Transmission}.
In particular, operating at cryogenic temperatures (\SI{\sim 4}{\kelvin}) and with strong pole-piece fields (\SI{\sim 2}{\tesla}) would enable spin polarizations exceeding \SI{0.3}{}, thereby greatly enhancing the signal ($\sim 10^4$ fold), compared to the \SI{2.3e-5}{}
 achieved in the present experiment.
The observed distance dependence suggests the ability to perform spatially selective measurements, potentially with resolutions down to the nanoscale \cite{ratzel2021controlling} as well as the possibility to drive higher-order multi-pole excitations, for example dipole forbidden electronic transitions which exhibit extended coherence times. This approach could serve as an alternative to generate time averaged painted potentials \cite{GAUTHIER2021, neumeier2024}.
While our demonstration focuses on a spin transition in the microwave frequency range, a class of transitions strongly utilized in quantum technologies due to its accessibility and long coherence times, similar results could be extended to the optical regime, enabling interaction with a broader class of quantum systems \cite{Gover2020free, Favro1971en, zhao2021quantum, ruimy2021toward, garcia2021optical,yalunin2021tailored, morimoto2021coherent}.
Our approach could evolve into a powerful tool for advancing the use of free-electron beams in coherent spectroscopic techniques \cite{ruimyFreeelectronQuantumOptics2025a}.

\section*{Acknowledgements}
We thank Wenzel Kersten for his fruitful comments on this manuscript; Johann Toyfl, Alexander Prüller for their help in the initial phase of the experiment, the USTEM and especially Thomas Schachinger for their expertise on SEM. Furthermore we thank our Workshop at Atominstitut with Heinz Matusch, Fabian Quurk and Walter Klikovich for their high quality work. PH and DR acknowledge the hospitality of the Erwin Schrödinger Institute in the framework of their "Research in Teams" project. PH thanks the Austrian Science Fund (FWF): P36041, P35953, Y1121. This project was supported by the ESQ-Discovery Program 2019 "Quantum Klystron (QUAK)" and the FFG project AQUTEM.



\onecolumngrid
\newpage
\appendix
\section{Data Analysis}
\label{sec:main_IQAppendix}

This section explains the isolation of the signal due to \emph{direct driving} by the near-field of the modulated electron beam in Figs.~\ref{fig:dataDiffEsr}\textbf{b-c} from the parasitic background and the indirect driving by the electromagnetic force. Finally, a qualitative analysis and a quantitative analysis are performed.

\subsection{Differential Measurement}
\label{sec:diffmeasurement}
The ESR signals induced by the near-field of the modulated electron beam are close to the thermal noise level of the selected lock-in amplifier bandwidth of \SI{0.53}{Hz}. 
Any stray microwave fields (e.g. from the microwave readout system) at approximately $\omega_0$ can generate parasitic background signals. To mitigate these and other background effects, we perform a differential measurement by recording two data sets: one with the electron beam on $f_\text{tot}\left(\omega\right)$ and a reference set with the beam off $f_\text{BG}\left(\omega\right)$, under otherwise identical conditions. Even though the modulation leads to a slight broadening of the signals, this broadening affects both the background and the beam signal equally. As a result, the background can still be subtracted from the total signal despite the non-linearity introduced by the modulation. The lock-in amplifier then detects the first harmonic at $\omega_m$, and due to the linearity of the Fourier transform the subtraction can be carried out explicitly.

\begin{figure}[htb]
    \centering
    \includegraphics{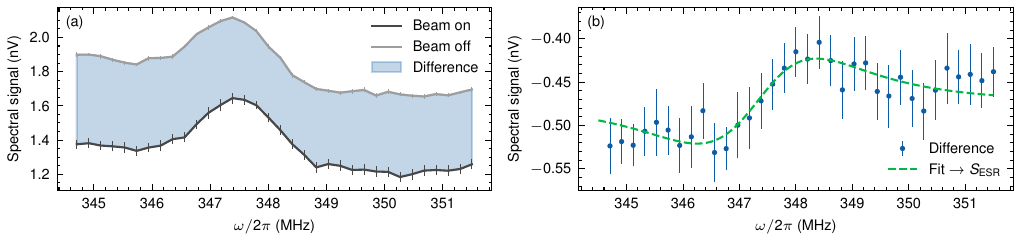}
    \caption{The differential measurement shown in Fig.~\ref{fig:dataDiffEsr}\textbf{a} effectively suppresses background contributions and isolates the signal induced by the near-field of the modulated electron beam. We acquire always two datasets, the electron beam switched on \textbf{a} and off \textbf{b}, yielding the difference in \textbf{c} which is fitted using Equ.~\ref{equ:esrfit} to extract signal size and phase for data analysis. The offset in \textbf{b} is caused by the field modulation, see Appendix~\ref{sec:calibration}.}
    \label{fig:diffscan}
\end{figure}

Fig.~\ref{fig:diffscan}\textbf{a} shows the two measurements, the difference is shown in \textbf{b}. This method allows to isolate the effect of the electron beam from the much larger background signal.

The parasitic ESR signal cannot be attributed to stray fields of our beam modulator, which is operated at the first harmonic $\frac{\omega_0}{2}$. Since our detection setup is only sensitive at the second harmonic $\omega_0$, and the driving signal is low-pass filtered, the fields radiated by the modulator at half the resonance frequency do not contribute to the background signal. It appears that the parasitic ESR signals originate from the local oscillator (LO, see Fig.~\ref{fig:rfschematic}) through an unidentified leakage or coupling mechanism. However, these background contributions are reliably suppressed by the use of differential measurements.

\subsection{Fitting Procedure}
\label{sec:diffMeasurementFitting}
To describe the shape of the ESR spectra, we recall that the solution of the steady-state Bloch equations in the rotating frame gives, with a small driving field with amplitude $B'_1$ at frequency $\omega$ and the resonance frequency $\omega_0 = - \gamma_e B_0$, the magnetization components \cite{weil_electron_2007}
\begin{equation}
    \label{equ:magn_y}
    M_{y'} = M_0\gamma_eB'_1 \frac{T_2}{1+(\omega-\omega_0)^2T_2^2}
\end{equation}
and
\begin{equation}
    \label{equ:magn_x}
    M_{x'} = M_0\gamma_eB'_1 \frac{\left(\omega-\omega_0\right)T_2^2}{1+(\omega-\omega_0)^2T_2^2},
\end{equation}
where $M_0 = \frac{n_e\gamma_e^2 \hbar^2 S(S+1)}{3k_BT_s}$ is the thermal magnetization with the spin density $n_e \sim \SI{1.5e27}{\m^{-3}}$ of the BDPA sample, the electron gyromagnetic ratio $\gamma_e$, the spin number $S=\frac{1}{2}$ and the sample temperature $T_S$. 
Depending on the phase of the LO, a linear combination $M_{y'}\cos\left(\phi\right) + M_{x'}\sin\left(\phi\right)$ is measured, where $\phi=n\pi,\ n \in \mathbb{Z}$ represents the absorptive case and $\phi=\frac{\pi}{2}+n \pi,\ n \in \mathbb{Z}$ the dispersive case. The lock-in amplifier modulated the resonance frequency $\omega$ with $A_m\cos\left(\omega_mt\right)$ and thus measures approximately the derivative.
We use the fit function (with $k=M_0\gamma_eB'_1$, $\Delta \omega = \omega - \omega_0$ and an additional offset parameter $d$)

\begin{equation}
    \label{equ:esrfit}
    f(\omega, \omega_0, \gamma_2, k, \phi, d) = k\left[\frac{d}{d\omega}\left(\frac{\gamma_2}{\gamma_2^2+(\Delta\omega)^2}\right)\cos(\phi) + \frac{d}{d\omega}\left(\frac{\Delta\omega}{\gamma_2^2+(\Delta\omega)^2}\right)\sin(\phi)\right]+d,
\end{equation}
where $\gamma_2 = \frac{1}{T_2} = \frac{\sqrt{3}}{2}\Delta\omega_{pp}$ and $\Delta\omega_{pp}$ is the peak-to-peak width of the derivative of the absorptive signal \cite{eatonQuantitativeEPR2010}.

We extract the signal magnitude from the peak-to-peak voltage of the fitted function $f\left(\omega, \omega_0, \gamma_2, k, \phi, d\right)$ (note that variations in $\gamma_2$ have an effect on the fitted $k$ value) and use it together with the phase $\phi$ to determine the two signal quadratures. For the small signals we acquired during the measurements we neglect the variations of $\omega_0$ and $\gamma_2$.
We attribute variations in the fit parameters $\omega_0$ and $\gamma_2$ to the reduced fit quality of the acquired small signals close to the noise floor and therefore neglect them.

Since we employ a lock-in amplification scheme with a modulation amplitude of $B_m \approx \SI{18}{\micro \T}$, the peak voltage of the detected signal (which is the first derivative) corresponds only to $\approx \SI{25}{\percent}$ of the on-resonance signal for absorptive shapes
in a hypothetical measurement scenario without a lock-in amplifier.

\subsection{Magnetic Field Created by the Microcoil Through Beam-Induced Electromotive Force}
\label{sec:coilexcitation}
The modulated beam-induced voltage in the microcoil $U_\text{EMF}$ causes a current flow limited by the ohmic resistance of the microcoil $R_c$  and the readout system. In the impedance-matched case, the resistance of the readout system is also $R_c$. Therefore, we approximate the current in the microcoil induced by the modulated beam close to resonance as $\frac{U_\text{EMF}}{2R_c}$ ($U_\text{EMF}$ refers to the voltage in the microcoil, the measured voltage is amplified by the gain of the microcoil impedance match and the readout system).
This current creates a magnetic field in $x$-direction that drives the spins with a $\pi/2$ phase shift relative to $B_1^x$ (sourced by the modulated electron beam), as expected from the time derivative of the law of induction. With $\omega_0 = 2 \pi\times\SI{348e6}{\per\s}$, a coil area $A_{coil} \sim \SI{1e-6}{\m^2}$ and two windings $N_{coil}=2$ one gets $U_\text{EMF} =\mathrm{i}N_{coil}\omega_0 A_{coil}\overline{B_1^x}$. The unitary magnetic field (the field normalized for \SI{1}{\A} current) of our microcoil $B_u \approx \SI{2e-3}{\T\per\A}$ is varying across the sample for different positions $\vec{r_i}$.  Since $B_u$ influences both excitation and detection, we use the self-weighted average $\frac{\sum_i B_u^2\left(\vec{r_i}\right)}{\sum_i B_u\left(\vec{r_i}\right)}$. The resulting parasitic driving field is approximated as
\begin{equation}
    B_p = I_{coil} B_u =\mathrm{i}\frac{N_{coil}\omega_0 A_{coil}}{2R_c} B_u \overline{B_1^x} \approx \mathrm{i} 3\overline{B_1^x}.
    \label{equ:EMF_BG}
\end{equation}

The ratio of the signals can be thus approximated as $S_\text{EMF}/S_\text{beam} \approx 3\mathrm{i}$ (see the ratio between $S_\text{ESR}(d=0)$ and $S_\text{ESR}(d\approx 0.7)$ in Fig. \ref{fig:main_IQ}\textbf{b}).
The resulting signal becomes (with $\alpha = R\frac{B_u}{2R_c}$)
\begin{equation}
    S_\text{EMF} = R \cdot \overline{B_1^x} = \alpha U_\text{EMF},
    \label{equ:S_EMF}
\end{equation}
where $R\approx \SI{4}{\V\per\T}$ is the ratio ESR signal to the driving field that can be determined using conventional ESR measurements (see Fig.~\ref{fig:bdpacrystal}\textbf{d}), involving a coarse estimate of the driving field. This gives a good estimate for $S_\text{EMF}$ and can be used to recover the direct driving signal $S_\text{beam}$.

\subsection{Phase Alignment}
\label{sec:phaseAlignment}
To retrieve the direct driving signal we have to subtract the estimate of the indirect driving signal $S_\text{EMF}$ from the measured ESR signal. As a prerequisite the phases of the two measurement devices (Lock-in amplifier measuring $S_\text{ESR}$ and the LTC5584 I-Q-demodulator measuring $U_\text{EMF}$, see Fig.~\ref{fig:rfschematic} for a more detailed schematic) have to be aligned. 
We achieve this by leveraging the constant offset in the ESR spectrum (see Fig.~\ref{fig:diffscan}\textbf{b}), which is caused by the modulation field $B_m$ (see Appendix~\ref{sec:calibration}) of the electron beam. 
The $U_\text{EMF}$ signal is mainly in one quadrature (except for spurious signals) and $\propto B_1^x$. By taking two ESR measurements in the $U_\text{EMF}$ maximum or mininum (see Fig.~\ref{fig:rfscan}) with a \SI{90}{\degree} relative phase shift and determining the offsets $d_I$ and $d_Q$, the phase shift of $S_\text{ESR}$ relative to $U_\text{EMF}$ is determined by $\arctan{\left(d_I/d_Q\right)}$.

\subsection{Qualitative Analysis}
\label{sec:qualitative}

\begin{figure} 
    \centering
    \includegraphics{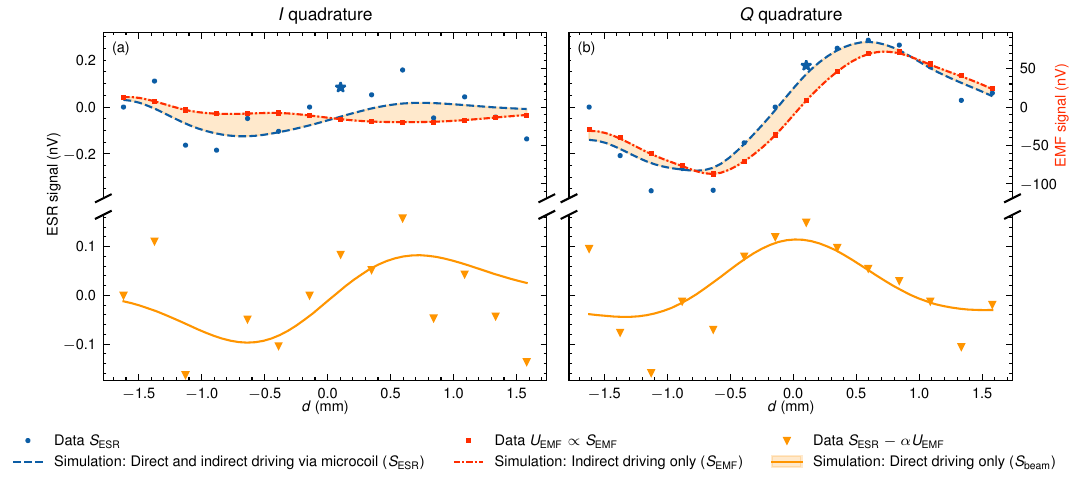}
    \caption{$I$ and $Q$ quadratures of the signals induced by a modulated electron beam at $h\approx\SI{0.45}{\mm}$ for different $d$. The indirect driving signal $S_\text{EMF}$ was estimated from the measured $U_\text{EMF}$ and then subtracted from the measured ESR signal $S_\text{ESR}$ to recover the direct driving signal $S_\text{beam}$. For the latter the quadratures correspond to the magnetization components driven by the $B_1^x$ and $-B_1^y$ field amplitudes, respectively. Due to the additional induction, the indirect driving signal is phase shifted by \SI{90}{\degree}, hence the $B_1^x$ field component induced into the microcoil appears in the $Q$ quadrature, which facilitates the disentanglement of the two effects (one has its maximum close to $d=0$, the other a zero-crossing).}
    \label{fig:main_IQ}
\end{figure}

Using Equ.~\ref{equ:EMF_BG} and~\ref{equ:S_EMF} we can recover the signal caused by direct driving of the electron beam's near-field. This can be simplified to a single factor 
$\alpha = R\frac{B_u}{R_c}$ ($R_c$ is the microcoil resistance, $B_u$ the coupling of the microcoil to the spins in \si{\T\per\A} and $R$ in \si{\V\per\T} the ESR signal to driving field ratio, determined from conventional ESR scans (see Fig.~\ref{fig:bdpacrystal}\textbf{d}) yielding:

\begin{equation}
    S_\text{beam} = S_\text{ESR} - \alpha U_\text{EMF}.
\end{equation}

The three components of this equation are shown as triangles, dots and squares in Fig.~\ref{fig:main_IQ} (which shows both signal quadratures for a beam at $h\approx\SI{0.45}{\mm}$ for different $d$, as in Fig.~\ref{fig:dataDiffEsr}\textbf{b}).
Due to the induction, $U_\text{EMF}$ exhibits an additional \SI{90}{\degree} phase shift and thus adds to the $B_1^y$ component of the near-field which exhibits a maximum close to $d=0$, as opposed to the former with the zero-crossing,  which simplifies the separation of direct and indirect driving fields (in the $Q$ quadrature). The $I$ quadrature comprises the $B_1^x$ component of the direct driving and residual induction into the microcoil. It should be noted that due to the uncertainy in the phase alignment (see above) the direct and indirect effect of the $B_1^x$ field component can mix between the two quadratures.

These results demonstrate that the direct driving signal can be recovered using a single scaling factor, leading to a clear observable shift of the measured ESR signal $S_\text{ESR}$ along $d$ with respect to the measured electromotive force in the microcoil ($U_\text{EMF}$).

\subsection{Quantitative Analysis Using Simulations}
\label{sec:quantitative}
In addition to the qualitative analysis performed above, the ESR measurements have been analyzed quantitatively to approximately validate the signal size (limited by uncertainties in the gain factors of the microcoil impedance match, the readout system and inaccuracies of the conventional ESR data). 
The solid lines in Fig.~\ref{fig:main_IQ} show the simulations, which are described in Appendix~\ref{sec:simulations}. A sample size of $1.1 \times 0.7 \times 0.7\, \si{\mm^3}$ was assumed and an estimate of the sample density was determined by conventional ESR measurement. The sample density and coil resistance were then adjusted for the simulations to fit the measured data, yielding $\approx 0.5n_e$ and $\approx \SI{1.25}{\ohm}$, respectively.
These are reasonable values, since the former varies between BDPA samples \cite{mitchellElectronSpinRelaxation2011} and the latter was estimated to $\approx \SI{1.75}{\ohm}$ using the impedance match and $\approx\SI{0.6}{\ohm}$ from conventional ESR saturation scans.
The data agree with the simulation, though the deviations are larger than those for the $I$ quadrature. This can be explained by the fact that the total signal $S_\text{ESR}$ is larger prior to subtraction.
Since spline interpolated $U_\text{EMF}$ data have been used for simulation of $S_\text{EMF}$, 
this explains the non-zero $S_\text{EMF}$ in Fig.~\ref{fig:main_IQ}\textbf{b}.

\subsection{Analysis Summary}
\label{sec:dataAnalysisConclusion}
We observe three signals which are different in size: (1) the parasitic ESR signal, which is subtracted by performing differential measurements, (2) $S_\text{EMF}$ the indirect driving with the electron beam via the microcoil, and finally (3) $S_\text{beam}$ the direct driving of spins by the near-field of the modulated electron beam. The last two can be disentangled because (2) experiences a \SI{90}{\degree} phase shift during the induction. The sum (2)+(3) exhibits a zero-crossing being shifted along $d$ with respect to the electromotive force induced in the microcoil. Finally, quantitative analysis has been used to confirm the measured signals.

\section{Experimental Setup}
\label{sec:details}
\textbf{Scanning electron microscope:} We utilize a Philips XL30 scanning electron microscope (SEM), equipped with a thermionic tungsten filament electron source, as our experimental platform. Two condenser lenses and various apertures collimate the beam.

\begin{wrapfigure}{r}{0.35\columnwidth}
    \includegraphics[width=0.35\columnwidth]{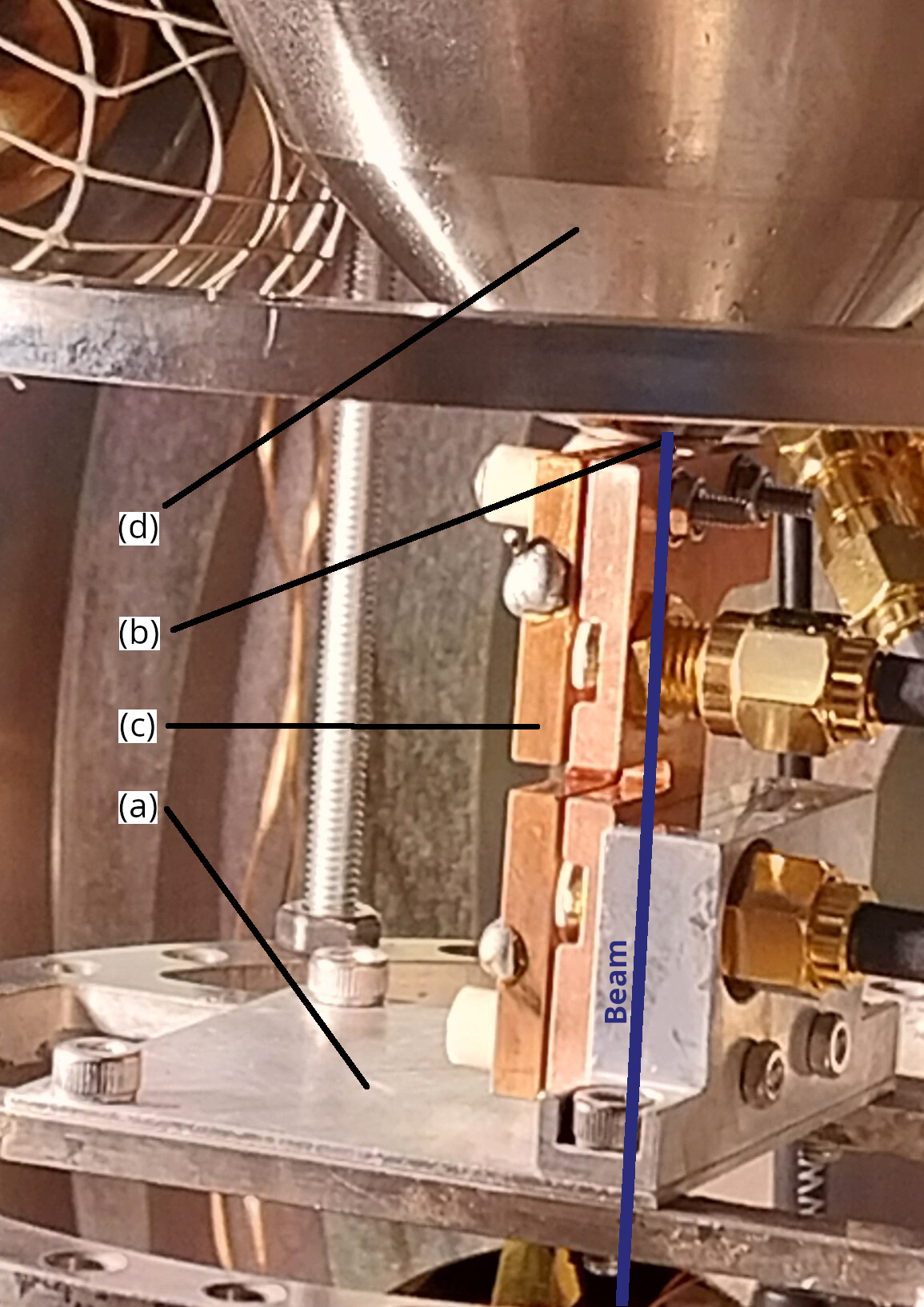}
    \caption{The \SI{4}{\mm} diameter \SI{40}{\mm} long custom deflection plates \textbf{c} are mounted on the SEMs pole shoe \textbf{d} in front of the final lens assembly and final aperture \textbf{b}. The aluminum plate \textbf{a} prevents beam distortions below the deflection plates.}
    \label{fig:kicker01}
\end{wrapfigure}

Since our experiment requires a sufficiently large magnetic near-field of the modulated electron beam we tried to maximize the beam current of the SEM while keeping basic imaging capabilities. We selected a \SI{600}{\micro \meter} final aperture and achieved currents in the range of \SIrange[range-phrase = -, range-units = single]{\sim 2.5}{5.5}{\micro \ampere} at \SI{10}{\kilo \electronvolt} ($v_e \approx 0.2 c$) in the sample region. We accomplished this by adjusting the position of the filament within the Wehnelt-cylinder and adjusting the bias voltages to project a larger portion of the filament tip towards the sample.

The non-standard operation regime yields unstable emissions. We attribute these to thermally induced filament deformation and deterioration that manifests both on a timescale of minutes, after enabling the beam and on a multiple hours timescale via slow drifts. In addition, the beam current exhibits a position dependence.
To compensate for those systematic errors, we measured the beam current via a copper foil on the bottom of the SEM chamber and rescaled our measurements accordingly.

\begin{figure*}
    \input{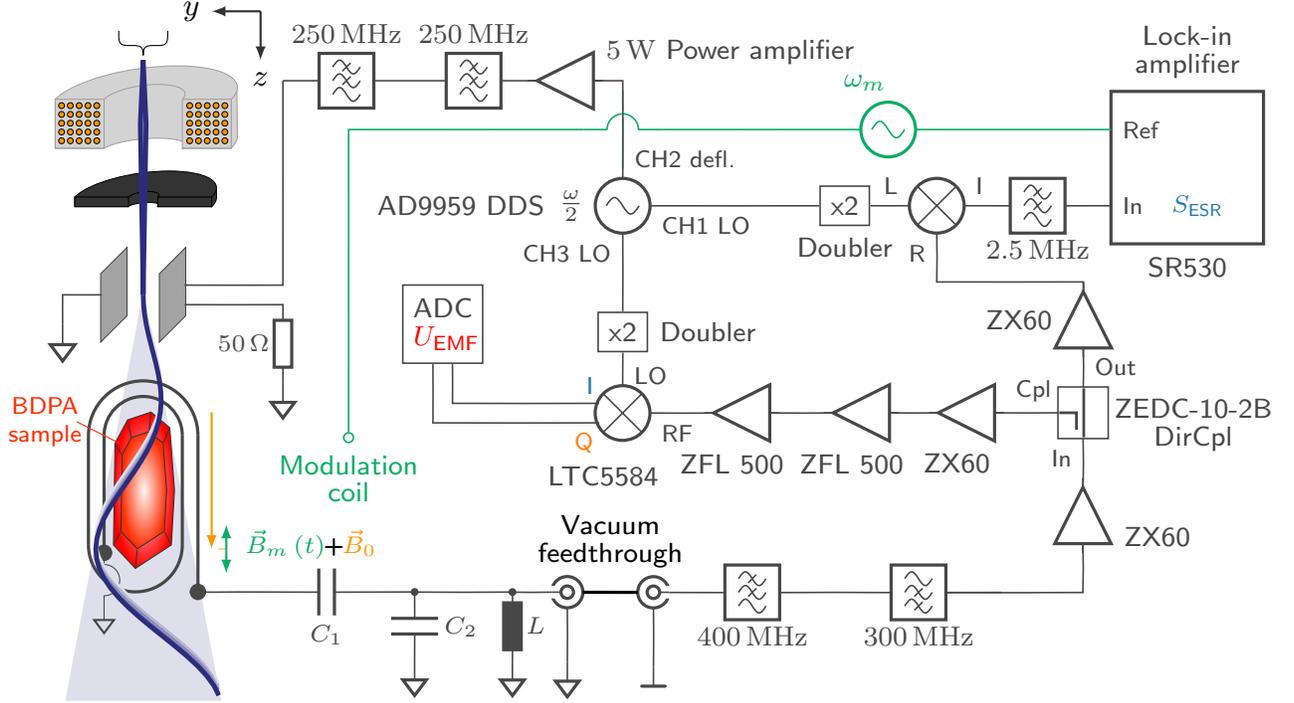}
    \caption{To measure the electromotive force (EMF) signal originating from the modulated electron beam induced spin precession, we use a lock-in detection scheme. The signal is inductively coupled to an impedance matched microcoil subsequently filtered and amplified. A directional coupler (DirCpl) splits it into: \textbf{CPL port} to directly monitor the near-field of the modulated electron beam and \textbf{Out port} for lock-in ESR measurements. In the excitation chain (CH2 defl), we utilize filters to prevent 2nd order and higher harmonics leaking from the beam modulator and thus parasitically driving the sample via radiated fields. The local oscillators (CH1 and CH3 of our DDS) are frequency doubled since our DDS only supports frequencies up to 250 MHz.}
    \label{fig:rfschematic}
\end{figure*}

\vspace{0.5em}
\label{sec:beamModulation}
\textbf{Beam Modulation:} For spatial electron beam modulation, we built a custom beam modulator (see Fig.~\ref{fig:kicker01}\textbf{c}) which consists of two half-cylindrical (radius \SI{2}{mm}) copper plates, each \SI{40}{mm} long, spaced \SI{500}{\micro \meter} apart. The modulator is mounted on a cage system that is fixed in front of the pole piece. A grounded aluminum plate (depicted in Fig.~\ref{fig:kicker01}\textbf{a}) shields against stray fields of the modulator assembly to further prevent deflection of the electron beam after it has passed the modulator. One of the deflection plates remains at ground potential and attaches directly to the pole piece, while the other one connects to a coaxial cable that delivers a microwave signal at half the readout frequency $\omega$, which is $\frac{\omega}{2}\approx 2\pi\times\SI{175}{\MHz}$. The microwave signal is then routed outside the SEM and terminated with a \SI{50}{\ohm} resistor to avoid thermal load inside the vacuum chamber of the microscope (see Fig.~\ref{fig:rfschematic}). Since the beam modulation occurs at around half the resonance frequency $\omega_0$, additional shielding to suppress direct microwave driving of the sample is not required.
This beam modulator allows for deflection amplitudes of up to \SI{2}{\mm} at the sample position \SI{130}{\mm} below the pole piece. 

\vspace{0.5em}
\label{sec:sampleStage}
\textbf{Sample stage:} The miniaturized ESR-setup is mounted on a custom stage, attached to a port aligner, in order to adjust the position of the ESR-sample relative to the electron beam.

\vspace{0.5em}
\label{sec:microwave_setup}
\textbf{Microwave setup:} The microwave setup is depicted in Fig.~\ref{fig:rfschematic}. The spin precession of our sample is inductively sensed by the microcoil (see Fig.~\ref{fig:bdpacrystal}\textbf{b}). This coil is impedance-matched to \SI{50}{\ohm} at $\approx \SI{350}{\mega \hertz}$ using an L-type capacitive matching network ($C_1=\SI{47}{\pico \farad}, C_2=\SI{18}{\pico \farad}$). A small inductance of $L=\SI{390}{\nano \henry}$ in parallel is used to protect our microwave electronics against low frequency high voltages. Subsequently, a low-pass (\SI{400}{\MHz}) and high-pass (\SI{300}{\MHz}) filter select the $2^\text{nd}$ harmonic components.  This step mitigates microwave background saturating the amplifiers. The filtered signal is amplified via one ZX60-P103LN+ low-noise amplifier. The signal is then split using a ZEDC-10-2B directional coupler in two paths: The \textbf{Out} port is amplified with a ZX60 low-noise amplifier. The signal is then extracted employing microwave down-mixing with a local oscillator (L) and low-pass filtered at \SI{2.5}{MHz}, to discard high-frequency sidebands. Subsequently a SRS-SR530 lock-in amplifier records the signal $S_\text{ESR}$. The \textbf{Cpl} port is amplified (1x ZX-60, 2x ZFL-500LN+) and used to monitor the two quadratures $I$ and $Q$ (see Fig.~\ref{fig:diffscan}\textbf{b}) of the signal $U_\text{EMF}$ induced by the electron beam via an LTC5584 I/Q demodulator, whose outputs are processed by two ADC channels of an RP2040 microcontroller. The latter signal is also used to estimate the electron beam position.

The microwave signal used to deflect the electron beam is sourced by a AD9959-based Direct Digital Synthesizer (DDS) (CH2, see Fig.~\ref{fig:rfschematic}) amplified by a \SI{5}{\watt} power amplifier and subsequently low pass filtered to suppress higher harmonics, thus preventing parasitic sample excitation by leaking fields.
Since the DDS supports frequencies only up to \SI{250}{\MHz}, CH1 and CH3 are passively frequency-doubled (FD-2+) and high pass filtered ($>\SI{300}{\MHz}$), and used as local oscillators (LO) in the I/Q-demodulator as well as the ESR mixer chain.

\section{Microcoil and Sample Preparation}\label{sec:pcb}
\begin{figure*}
    \centering
    \includegraphics[width=0.42\textwidth]{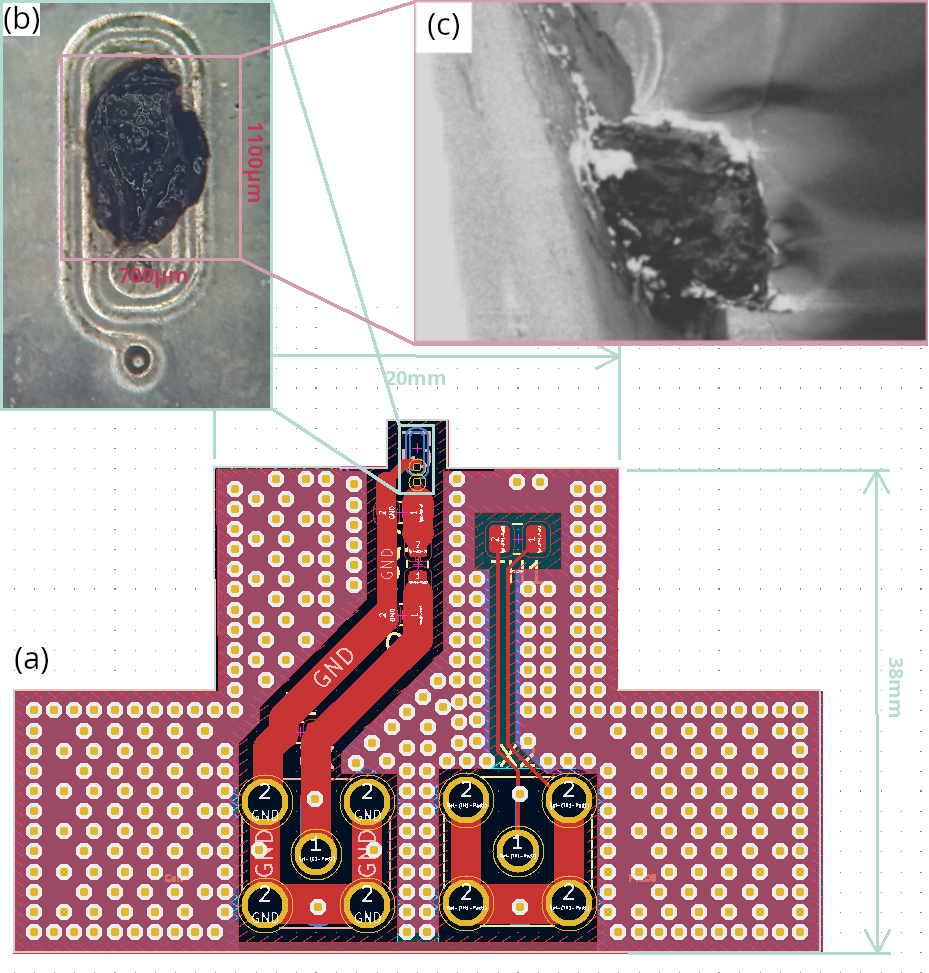}
    \includegraphics{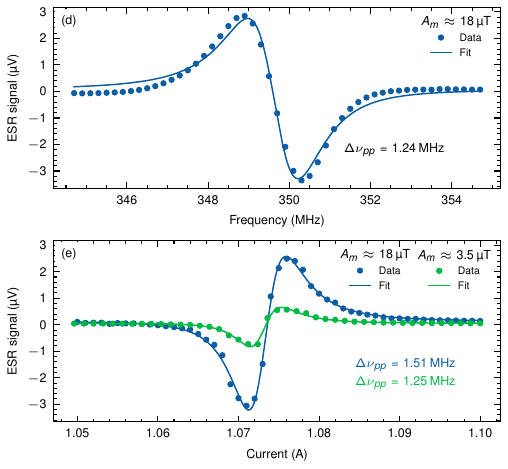}
    \caption{\textbf{a} Detailed view of the PCB layout in the design software. The vias spread over the surface are used for thermal management and contact to our stage ground. \textbf{b} Top view of the PCB with a BDPA sample (which was used in another measurement) inside the microcoil. \textbf{c} SEM image of the sample. The curved lines to the top of the sample  are caused by charging effects on the non conductive surface of the sample and due to the fact that the beam is blocked on a wrinkled conductive foil used as Faraday cup $\SI{35}{mm}$ below the sample. The geometry appears slightly curved due to the extension of the PCB and sample along the beam-axis as well as the applied $B_0$ field which causes a position dependent twist of the image plane.
    \textbf{d}, \textbf{e} show conventional ESR spectra employing a directional coupler to excite and sense the sample via the microcoil by performing \textbf{d} a frequency sweep and \textbf{e} a current sweep. The modulation amplitude applied during the measurements involving the electron beam ($A_m \approx \SI{18}{\micro\T}$) causes a slight broadening (\textbf{e}). The fitted signal in \textbf{d} shows a narrow absorption linewidth, which may be attributed to the frequency dependence of the microwave setup.}
    \label{fig:bdpacrystal}
\end{figure*}

The microcoil sensing for ESR signals is fabricated on a printed circuit board (PCB, depicted in Fig.~\ref{fig:bdpacrystal}\textbf{a}).
The PCB uses a standard \SI{1.6}{mm} FR-4 core with \SI{35}{\micro\metre} copper layers on both sides.
The microcoil (see Fig.~\ref{fig:bdpacrystal}\textbf{b}) is printed on the opposite side of the matching network (see Fig.~\ref{fig:bdpacrystal}\textbf{a}). This design prevents the electron beam hitting the matching network and creating parasitic signals and noise in the readout circuit.

For the spin ensemble in our experiment, we used a BDPA crystal $1.1 \times 0.7 \times 0.7\,\si{\mm^3}$, which is placed atop a groove within the microcoil and fixed with small amounts of vacuum grease (see Fig.~\ref{fig:bdpacrystal}\textbf{b}). The single crystal extends \SI{\sim 700}{\micro \m} above the PCB surface.

We estimated the microcoil resistance $R_c$ to be on the order of \SI{1}{\ohm} at $\omega_0\approx2\pi\times\SI{350}{\MHz}$.

All copper layers are grounded to discharge unwanted surface charges deposited by the electron beam. To further reduce beam distortion by charging effects, the surface of the PCB including the microcoil was coated with a \SI{\sim 10}{\nano \meter} thick 60:40 gold-palladium layer on top of an insulating solder mask.

To evaluate the Q-factor of the combined coil and impedance matching setup, we measured the resonance depth and bandwidth. At \SI{350}{MHz} the Q-factor was determined to be approximately 100 with a resonance depth of \SI{22}{dB} and a bandwidth of \SI{3}{MHz}.

We first characterized the sample and measurement setup by performing conventional ESR measurements, by exciting the sample directly with the microcoil with a driving field injected via a directional coupler (DirCpl) inserted directly at the vacuum feedthrough (the driving field then amounts to a few \si{\micro\T}, i.e. below saturation which would occur at $\sim \SI{100}{\micro\T}$ for BDPA) and measuring the spin precession at the coupling port of the DirCpl. A spectrum recorded this way, with a frequency sweep, is shown in Fig.~\ref{fig:bdpacrystal}\textbf{d} for a modulation amplitude $A_m \approx \SI{18}{\micro\T}$. Fig.~\ref{fig:bdpacrystal}\textbf{e} depicts the same spectrum with a $B_0$ sweep at constant frequency $\omega_0 = 2\pi\times\SI{349.7}{\MHz}$. The smaller linewidth in the frequency sweep (at $A_m \approx \SI{18}{\micro\T}$) is most likely explained by the convolution with the narrow impedance match of the microcoil. Lock-in detection with a modulated $B_m$-field broadens the linewidth from $\Delta\omega_{pp} \approx 2\pi\times \SI{1.25}{\MHz}$ at low $A_m$ to $\approx 2\pi\times \SI{1.5}{\MHz}$ for $B_m = \SI{18}{\micro\tesla}$ (see Fig.~\ref{fig:bdpacrystal}\textbf{e}). From the current sweep at low $A_m\approx \SI{3.5}{\micro\T}$ we determined a transversal relaxation time $T_2 \approx \SI{147}{\ns}$, consistent with previously reported values \cite{PhysRevLett.4.13,mitchellElectronSpinRelaxation2011}.

\section{Phase Calibration}
\label{sec:calibration}
\begin{figure}[htb]
    \centering
    \includegraphics{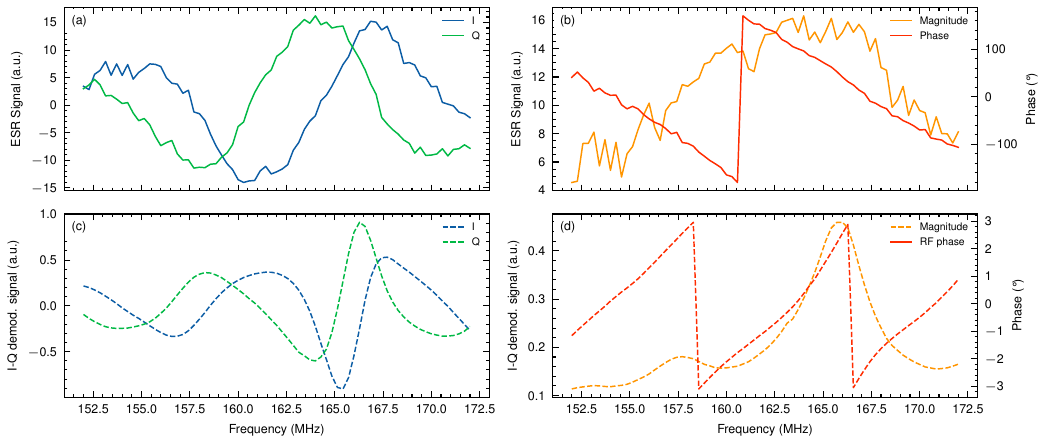}
    \caption{Due to different cable lengths and other phase shifting elements in our setup, the phase relation between the exciting microwave input and reference LO input at our mixer (or I-Q-demodulator in case of the electron beam-induce EMF) changes with frequency $\omega$. 
    \textbf{a}, \textbf{b} for our ESR scans, we only measure one quadrature at a time. I and Q refer to two different measurements with \SI{90}{\degree} phase difference. The signal in \textbf{a} is a residual electron beam EMF caused by the modulation coil, which slightly affects the beam. We use the linear phase dependence in \textbf{b} to shift the LO, thereby removing the frequency dependence and always maintain the same relative phase. A signal offset (which still varies with the LO phase) remains (see Fig.~\ref{fig:diffscan}\textbf{b}).
    \textbf{c}, \textbf{d} In case of the electron beam EMF, both quadratures are recorded simultaneously and can be used to determine magnitude and phase. The impedance match of the microcoil can be seen in the EMF magnitude (note that the measurements in this figure were taken with a different PCB).}
    \label{fig:phasecalibration}
\end{figure}

The modulated electron beam also causes an additional background in the ESR spectrum, shown in Fig.~\ref{fig:phasecalibration}\textbf{a} - the $I$ and $Q$ data were taken with two measurements with a \SI{90}{\degree} phase shift of the LO. The modulation for the lock-in detection also shifts the beam position due to the Lorentz force inducing an EMF at the lock-in modulation frequency in the microcoil, which can be detected by the lock-in amplifier. This is expected to give a constant offset in the signal, limited by the impedance match of the microcoil. However, since we perform frequency sweeps, different microwave path lengths from the frequency source (DDS) to the \emph{R} port and the \emph{L} port of the mixer (see Fig.~\ref{fig:rfschematic}) cause the signal to move between the quadratures, explaining the observed behavior. To fit this behavior, we use a function $g$, which is a  combination of a cosine and the impedance match
\begin{equation}
    g(\omega, k, a, \theta, \omega_0, s) = k\cos(a\omega+\theta)\cdot \frac{1}{\sqrt{\left(\frac{\omega-\omega_0}{s}\right)^2+1}},
\end{equation}
where $k\cos\left(a\omega+\theta\right)$ describes the frequency-dependent shift of the signal in and out the quadrature with factor $a$ (unit is \si{\s}) for the frequency $\omega$, to get the "frequency" of the phase shift and a phase $\phi$, and $\frac{1}{\sqrt{\left(\omega-\omega_0\right)^2/s^2+1}}$ fits the impedance match of the coil with resonance frequency $\omega_0$ and width $s$.
In Fig.~\ref{fig:phasecalibration}\textbf{b}, magnitude and phase are plotted.
By adding a frequency dependent phase to the LO, one can compensate for the phase shift, yielding a nearly flat background, only affected by the frequency dependence of our microwave setup (see Fig.~\ref{fig:diffscan}\textbf{b}).
Fig.~\ref{fig:phasecalibration}\textbf{c}, \textbf{d} show the frequency dependence of the EMF of the electron beam. 

\section{Finite-Element Simulation}
\label{sec:simulations}
To model the ESR signals, a finite element simulation of the sample was carried out for the magnetic near-field $B$ of the modulated electron beam driving the sample.
We assume a cuboid sample geometry of $1.1 \times 0.7 \times 0.7 \,\si{\mm \cubed}$, inferred from microscopic images (see Fig.~\ref{fig:bdpacrystal}\textbf{b}).
The spin density was first estimated to $\approx 0.25 n_e$, using the signal amplitude observed in the conventional ESR measurement shown in Fig.~\ref{fig:bdpacrystal}\textbf{d} and later adjusted to $\approx 0.5n_e$ to match the data involving the electron beam. This deviation is acceptable when considering errors in gain of the impedance match, the setup inside the vacuum chamber, of our readout setup and the fact that the driving fields in conventional ESR are not precisely known. 

The simulation was performed on a rectangular grid composed of voxels with a side length of \SI{100}{\micro\m}. 
The results reveal a pronounced spatial dependence of the driving field, arising from the near-field of the electron beam and its position relative to the sample (shown as solid lines in Fig.~\ref{fig:dataDiffEsr}\textbf{b}, and as orange solid lines in Figs.~\ref{fig:main_IQ} and~\ref{fig:spatres_verbose}).

\subsection{Simulation of the near-field}
\label{sec:siulationNearField}

\begin{wrapfigure}{r}{0.23\columnwidth}
    \centering
    \vspace{-25pt}
    \includegraphics{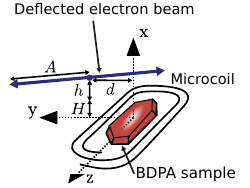}
    \caption{Sketch of the electron beam deflection in the $xy$ plane (only 2D representation) with respect to the spin sample and the microcoil, including the tilt and the beam position $d$ and $h$ (the origin is in the center of the microcoil, $h$ is the distance from the top of the sample with height $H=\SI{0.7}{\mm}$. }
    \vspace{-60pt}
    \label{fig:sketch}
\end{wrapfigure}
The electron beam is modeled as an infinitely long thin beam at position $(x, y) = (h, d)$ that is moving side-by-side of the observer with frequency $\omega/2$ at position $\mathbf{r} = \begin{pmatrix}
    x, y, 0
\end{pmatrix}$. Using Ampère's law the time dependent magnetic field becomes (using \cite{ratzel2021controlling}, Eq.~G3)
\begin{equation}
    \mathbf{B}\left(\mathbf{r}, t\right) = -\frac{\mu_0I}{2\pi} \frac{\begin{pmatrix}y-d-A\cos\left(\frac{\omega}{2}t\right)\\h-x\\0\end{pmatrix}}{\left(y-d-A\cos\left(\frac{\omega}{2}t\right)\right)^2+\left(h-x\right)^2}.
    \label{equ:nearfield}
\end{equation}
The harmonic components are then determined numerically by calculating the field for different $t$ and performing a Fourier transform.
Fig.~\ref{fig:sketch} shows the beam modulation (sketched only in the $xy$ plane) relative to the sample. The \SI{5}{\degree} angle of the deflection direction with respect to the $y$-axis is also considered in the simulations, leading to a slight modification of Eq.~\ref{equ:nearfield}.
In addition, finite element simulations have been conducted using Biot Savart's law, showing that the approximation of the beam as infinite line agrees unless $A \gg h$, which is sufficient for this experiment.

\subsection{Bloch Equations}
\label{sec:blochEqu}
The near-field of the modulated electron beam exhibits two field components, hence the Bloch equations (\cite{weil_electron_2007}, Eqs.~10.14 and~10.17; $T_1$ and $T_2$ are the longitudinal and transverse relaxation times, respectively)
\begin{equation}
    \frac{\mathrm{d}\mathbf{M}}{\mathrm{d}t} = \gamma_e \mathbf{M}\times \mathbf{\widetilde{B}} - \frac{1}{T_1}\begin{pmatrix}0\\0\\M_z-M_0\end{pmatrix} - \frac{1}{T_2}\begin{pmatrix}M_x\\M_y\\0\end{pmatrix}
\end{equation}
have to be solved for two driving field components with amplitudes $B_1^x$ and $B_1^y$ and phases $\theta_x$ and $\theta_y$ to be included in the simulations.
For
\begin{equation}
    \mathbf{\widetilde{B}} = \begin{pmatrix}B_1^x\cos(\omega t+\theta_x)\\B_1^y\cos(\omega t+\theta_y)\\B_0\end{pmatrix}
\end{equation}
we can transform the equations into the rotating frame using
\begin{equation}
    R_z\left(-\omega t\right) = \begin{pmatrix}
        \cos\left(\omega t\right) & -\sin\left(\omega t\right) & 0 \\
        \sin\left(\omega t\right) & \cos\left(\omega t\right) & 0 \\
        0 & 0 & 1\\
    \end{pmatrix}.
\end{equation}
Solving the equations for low driving powers leads to the magnetization components
\begin{equation}
    M_{x'} = -\frac{M_0\gamma_e T_2}{1+\left(\Delta\omega T_2\right)^2} \left(B_1^{y'} - T_2\Delta\omega B_1^{x'} \right)
\end{equation}
and
\begin{equation}
    M_{y'} = \frac{M_0\gamma_e T_2}{1+\left(\Delta\omega T_2\right)^2} \left(B_1^{x'}+T_2\Delta\omega B_1^{y'}\right),
\end{equation}
 where $B_1^{x'} = \frac{1}{2}\left(B_1^x\cos\theta_x - B_1^y\sin\theta_y\right)$ and $B_1^{y'} = \frac{1}{2}\left(B_1^x\sin\theta_x + B_1^y\cos\theta_y\right)$. 

Transforming back from the rotating frame (using $R_z\left(\omega t\right)$ the magnetization components become
\begin{equation}
        \begin{split}
                M_x & = \mathrm{Re}\left[\left(M_{x'} - \mathrm{i} M_{y'}\right)e^{\mathrm{i} \omega t}\right] \\
                M_y & = \mathrm{Re}\left[\mathrm{i}\left(M_{x'} - \mathrm{i} M_{y'}\right)e^{\mathrm{i} \omega t}\right]. \\
        \end{split}
\end{equation}
The spin-induced EMF signal can then be calculated using the principle of reciprocity \cite{hoult1969}. We define $\mathbf{B}_u=(B_u^x,B_u^y,B_u^z)$ as the magnetic field at the position of the sample that is generated by a \SI{1}{\A} current through the microcoil with units T/A. Then, equation (2) of \cite{hoult1969} implies for the voltage that is induced in the microcoil due to the magnetization of the sample (with the voxel volume $V_v$, $\Arg$ denotes the argument of a complex number)
\begin{equation}
    U = -V_v\frac{d}{dt}  \mathbf{B}_u\cdot \mathbf{M} = 
    \mathrm{Re}\left[ i\omega V_v \left(B_u^x + \mathrm{i} B_u^y\right) \left(M_{x'}-\mathrm{i}M_{y'}\right)e^{i\omega t} \right]
    = 2\mathrm{Re}\left[U_0 e^{i\omega t}\right]
    = \left|U_0\right|\cos\left(\omega t+ \Arg U_0\right)
\end{equation}
with
\begin{equation}
    \label{equ:esremf}
    U_0 = \frac{\mathrm{i}}{2}\omega V_v \left(B_u^x + \mathrm{i} B_u^y\right) \left(M_{x'}-\mathrm{i}M_{y'}\right) \propto B_1^xe^{\mathrm{i}\theta_x}-\mathrm{i}B_1^ye^{\mathrm{i}\theta_y}.
\end{equation}
Therefore, magnitude and phase of the signal $U$ depend on the interference of the $x$ and $y$ field components in the rotating frame and we can regard $B_1^xe^{\mathrm{i}\theta_x}-\mathrm{i}B_1^ye^{\mathrm{i}\theta_y}$ as the driving field of the spins.
Since $\omega_0 = -\gamma_e B_0$ the spins precess around the negative $z$ axis which explains the minus sign for $B_1^y$.

In the finite-element simulations, the signal induced into the microcoil is calculated as the sum of the complex $U_0$ which are calculated for each voxel using Equ.~\ref{equ:esremf}.

\subsection{Indirect Driving}
\label{sec:indirectDriving}
The simulations for data analysis are based on the spline interpolation of the measured $U_\text{EMF}$ values (see Fig.~\ref{fig:rfscan}). The corresponding driving field is calculated for every finite element, (see Appendix~\ref{sec:simulations}), as

\begin{equation}
    B_\text{EMF}\left(\vec{r_i}\right) = \left(B_u^x\left(\vec{r_i}\right)+\mathrm{i}B_u^y\left(\vec{r_i}\right)\right)\frac{U_\text{EMF}}{2R_c}.
    \label{equ:sim_EMF_BG}
\end{equation}

\begin{figure} 
    \centering
    \includegraphics{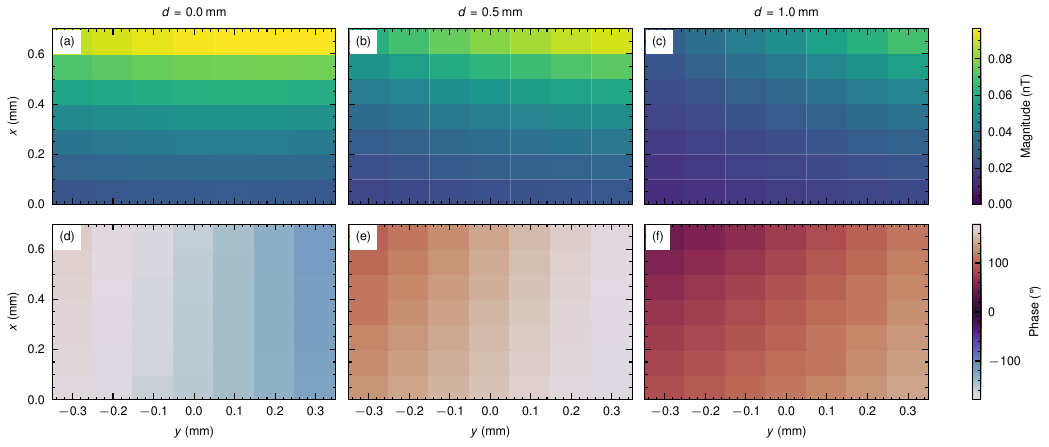}
    \caption{Spatial dependence of the complex magnetic field $B_1^x-\mathrm{i}B_1^y$ of the modulated electron beam at \SI{1}{\micro\ampere}, as experienced by the sample (voxels with $\SI{100}{\micro\m}$ side length), averaged along the $z$-axis. The origin of the coordinate system is found in the center of the microcoil, see Fig.~\ref{fig:sketch}. \textbf{a}-\textbf{c} depict the magnetic field magnitudes for different electron beam positions along the $y$-axis ($h=\SI{0.6}{\mm}$) with a deflection amplitude $A=\SI{0.9}{\mm}$ approximately along the $y$-axis (tilted by \SI{5}{\degree} into the $xy$-plane) nearby the sample. \textbf{d}-\textbf{f} display the corresponding phases, which also exhibit a significant spatial dependence.}
    \label{fig:paintedpotential}
\end{figure}

\subsection{Variation of the driving field across the sample}
Fig.~\ref{fig:paintedpotential} shows the simulation of the complex magnetic field $B_1^x-\mathrm{i}B_1^y$ experienced by different voxels at various electron beam positions along the $y$-axis ($h=\SI{0.6}{\mm}$), averaged along the symmetric $z$-axis (beam direction).

For the simulation, we assumed a beam deflection with an amplitude of \SI{0.9}{\mm} along the $y$-axis for several beam positions $d$ at $h=\SI{0.6}{\mm}$ under an angle of \SI{5}{\degree} relative to the $y$-axis, as sketched in Fig.~\ref{fig:sketch}. 
Fig.~\ref{fig:paintedpotential}\textbf{a}-\textbf{c} illustrate how the field magnitude at the sample position decreases with distance, while Fig.~\ref{fig:paintedpotential}\textbf{d}-\textbf{f} shows the spatially varying phase of the driving field. 

This spatial dependence could be used in future applications to create painted potentials or to enhance spatial resolution.

\section{Near-Field Induced Electromotive Force}
\label{sec:nearFieldEMF}
During the ESR measurements, the electromotive force ($U_\text{EMF}$) due to the modulated electron beam in the microcoil is simultaneously recorded (see Appendix \ref{sec:microwave_setup}). 
The differential measurement allows to recover the field of the electron beam $b\cos(\omega t+\phi_2)$ by assuming that we measure the superposition $C\cos(\omega t+\phi) = a\cos(\omega t + \phi_1) + b\cos(\omega t+\phi_2)$ when the electron beam is on and the signal background $a\cos(\omega t+\phi_1)$ when the electron beam is off. 
In the complex representation we get $Ce^{i\phi}-ae^{i\phi_1} = be^{i\phi_2}$.

\begin{figure}[htb]
    \centering
    \includegraphics{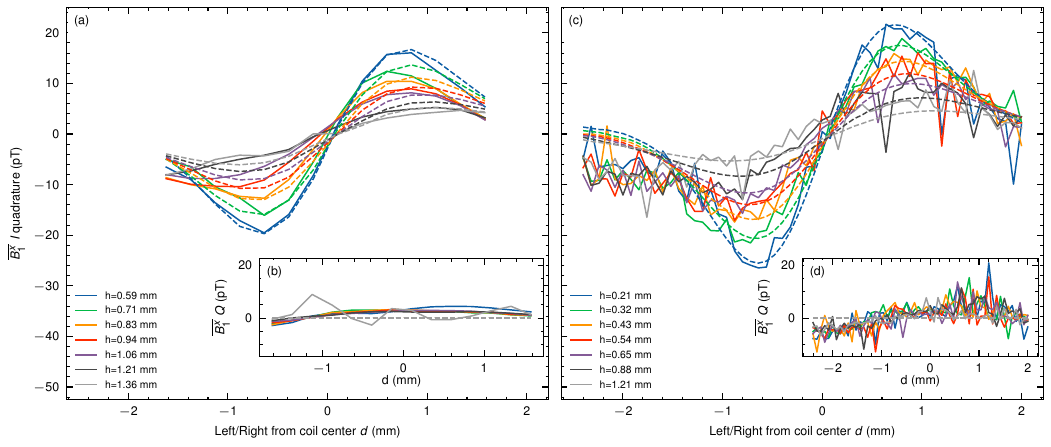}
    \caption{Second harmonic Fourier component of the magnetic near-field of the modulated electron beam in $x$-direction ($\overline{B_1^x}$) inferred using the electromotive force (EMF) induced in the microcoil normalized to \SI{1}{\micro\ampere} beam current. Two quadratures $I$ and $Q$ of the detected signal are shown in \textbf{a,c} and \textbf{b,d}, respectively. \textbf{a}, \textbf{b}: The data was recorded at the same time as the data in Fig.~\ref{fig:spatres_verbose}. With an approximate beam deflection amplitude of \SI{0.9}{\mm} nearby the sample at $\omega = 2\pi\times \SI{174}{\MHz}$ we determined scaling and distances comparing it to simulations. The scaling along the $x$-axis had to be corrected by a factor $0.5$. \textbf{c}, \textbf{d}: Fast data recording at $\omega = 2\pi\times\SI{174}{\MHz}$: No ESR spectra were taken simultaneously. The data are in agreement with the position scaling determined from from SEM images.}
    \label{fig:rfscan}
\end{figure}

The magnetic field amplitude $\overline{B_1^x}$ in $x$-direction, averaged over the coil area, is calculated from the modulated beam induced voltage $U_\text{EMF}$, which is extracted from the differential measurement (see Appendix~\ref{sec:main_IQAppendix}),
using
\begin{equation}
    \overline{B_1^x} = \frac{U_\text{EMF}}{\omega N_\text{coil} A_\text{coil}},
\end{equation}
where $A_\text{coil}$  is the coil area and $N_\text{coil}$ the number of coil windings.
$\overline{B_1^x}$ is used to determine the position of the microcoil relative to the electron beam (the scaling along the $x$-axis and $y$-axis as well as the deflection amplitude have been estimated using the imaging capabilities of the SEM).
Fig.~\ref{fig:rfscan}\textbf{c}, \textbf{d} show a $\overline{B_1^x}$ measurement at $\omega = 2\pi\times \SI{174}{\MHz}$ (\SI{\sim 5}{\s} per position and no ESR spectra recorded). In contrast, the data from the ESR measurements (\SI{\sim 20}{\min} per position) shown in Fig.~\ref{fig:rfscan}\textbf{a}, \textbf{b} reveal a slightly different behavior along the $x$-axis (moving away from the sample), which we attribute to charging and/or drifts of the electron beam position. 
To account for these, we rescale the $x$-axis accordingly.
The asymmetry between the two peaks in \textbf{a} is attributed to the small angle of \SI{5}{\degree} between the deflection- and $y$-axis (see Fig.~\ref{fig:sketch}).

As predicted by theory, $\overline{B_1^x}$ can be aligned to fit in one quadrature ($I$, in Fig.~\ref{fig:rfscan}\textbf{a}, \textbf{c}). However, some smaller residuals are found in the $Q$ quadrature (Fig.~\ref{fig:rfscan}\textbf{b}, \textbf{d}).

\section{Resonance Shift from Electron Beam's DC Component}
\label{sec:resonanceShift}
For a \SI{4}{\micro\ampere} beam at a distance of \SI{100}{\micro \meter} from the sample, our numerical simulations estimate the DC component of the magnetic field to be below \SI{10}{\nano\T} (see Section~\ref{sec:simulations}).
With a resonance frequency of \SI{\sim 28}{\GHz\per\T}, this limits the maximum shift of the ESR resonance frequency to \SI{280}{\Hz} and should therefore, not cause any significant line shift that could resemble the shape of a signal when performing differential measurements.

\section{Detailed Spatial Dependence}
\label{sec:verbosePlots}
\begin{figure}[htb]
    \centering
    \includegraphics{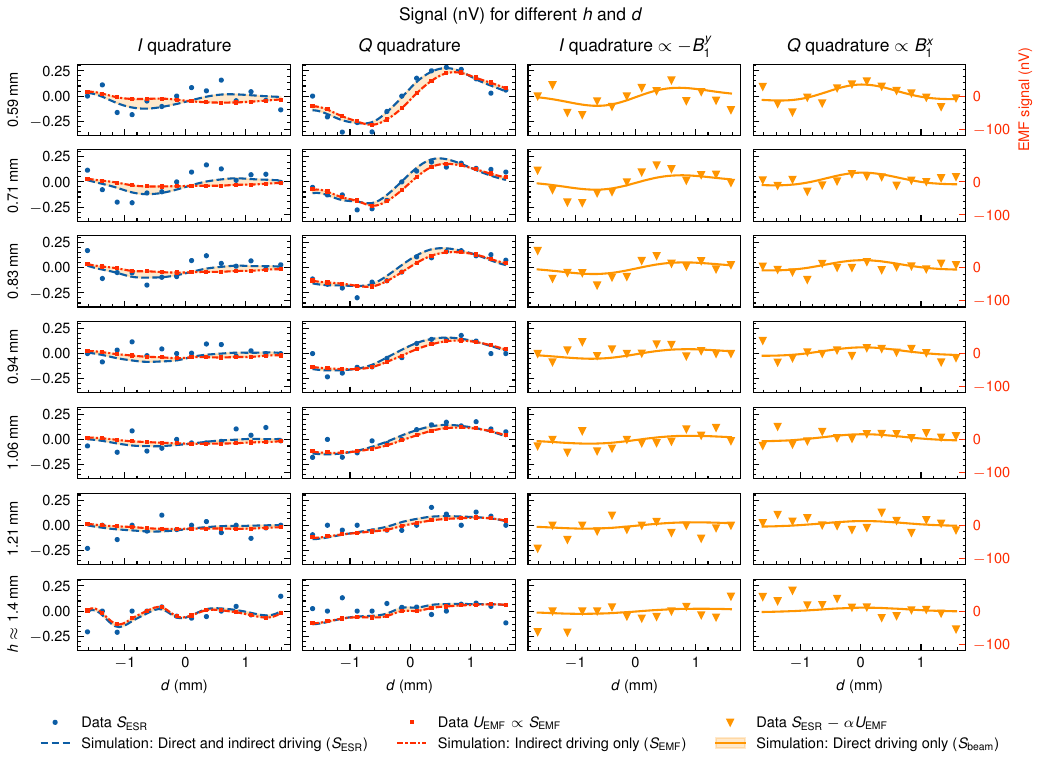}
    \caption{Overview of all fitted data points used in Fig.~\ref{fig:dataDiffEsr}\textbf{c}. Each line shows the data for a certain $h$, the two left columns show the $S_\text{ESR}$, $U_\text{EMF}$ and $S_\text{EMF}$ signal compared to simulations. The two right columns depict the two quadratures of the recovered $S_\text{beam}$ signal that is directly driven by the modulated electron beam, again comparing to simulations. The measurement conditions are the same as the measurement shown in Fig.~\ref{fig:dataDiffEsr}\textbf{b}, except for the different $h$.}
    \label{fig:spatres_verbose}
\end{figure}

Detailed scans for different $(h, d)$ are shown in Fig.~\ref{fig:spatres_verbose}, exhibiting the signal driven by the two field components $B_1^x$ and $B_1^y$ along $d$ and a decay along $h$. While the total $S_\text{ESR}$ and indirect driving signal $S_\text{EMF}$ depicted in the second column is clearly visible also for larger distances $h$ the smaller direct driving signal $S_\text{beam}$ on the third and fourth columns vanishes quickly.

\end{document}